# Variant effects on protein-protein interactions: methods, models and diseases

Sven Larsen-Ledet[1], Aleksandra Panfilova[1] and Amelie Stein[1,*]

1: Department of Biology, University of Copenhagen, Ole Maaløes Vej 5, DK-2200 Copenhagen, Denmark.

*: to whom correspondence should be addressed amelie.stein@bio.ku.dk

**Abstract**

Advances in sequencing have revealed that each individual carries about 10,000 missense variants. For the vast majority, we do not know what the functional consequences - if any - will be. Further, mechanistic insight, such as structural details, would be immensely helpful in development of therapeutic approaches. Here we review recent developments in experimental and computational techniques aimed to assess the impact of variants on protein-protein interactions, including limitations and upcoming challenges.

**Introduction**

Proteins are essential molecules within the cell, they facilitate most cellular processes, often through interaction with other proteins. Their functionality often depends on their amino acid sequence and structure. Changes to this sequence, even minor changes in the form of single amino acid substitutions (also termed missense variants[1]) can perturb protein function and potentially lead to diseases. Recent advances in human genome sequencing have identified millions of such missense variants in protein-coding genes, some linked to monogenic diseases [1,2]. However, many of these variants are extremely rare in the population and classified as variants of uncertain clinical significance [3,4]. It remains a major challenge to assess not only the functional consequences of these variants but also understand the underlying molecular mechanisms by which they cause diseases. Addressing this issue could enhance clinical diagnostics and patient counselling and pave the way for precision medicine to provide more effective treatments tailored to individual patients [5].

---

[1] In the context of human genomics, the term "variant" is used for differences from the reference sequence. In biophysics, sequence changes are usually referred to as "mutations", partly due to the fact that many of these originally were engineered rather than observed in individuals. As we here cover work from both areas, we have decided to use the more appropriate term in the respective section.

Protein interactions with other proteins and biomolecules (DNA, RNA, lipids, or small molecules) mediate virtually all cellular processes, including signal transduction pathways, protein degradation, cell cycle regulation, gene expression, and DNA repair. Here, we define protein-protein interactions (PPIs) as the physical interaction between two or more proteins, which can vary in complexity, affinity, and heterogeneity, ranging from homo-/heterodimers to multisubunit complexes (**Fig. 1**). The numerous PPIs in the cell are organized into tightly regulated networks known as the interactome [6]. These networks facilitate precise spatiotemporal regulation of cellular processes in response to internal and external stimuli [7]. The interactome is often depicted as a network of proteins (nodes), with edges representing the observed interactions. Individual proteins may participate in various interactions depending on their subcellular localization, expression levels, and post-translational modifications [8]. Distinct protein variants in the same gene may affect none (quasi-wild-type), some (edgetic), or all (quasi-null) of the interactions, leading to different interactome profiles and phenotypic effects [9].

Quasi-null variants often cause destabilization and degradation, resulting in complete loss of the protein and all its interactions. Most proteins must adopt a stable, folded structure to be functional. Thus, protein stability is a fundamental requirement for protein function. Protein stability refers to the thermodynamic stability of a protein [10], however, protein abundance in the cellular environment is influenced not only by thermodynamic stability but also by factors such as macromolecular crowding [11], interactions with other molecules [12,13], subcellular localization [14], post-translational modifications [15] and changes in gene expression [16]. Despite this, it has been demonstrated that protein stability and protein abundance are closely related for several proteins [10–13]. Consequently, missense variants can lead to loss-of-function phenotypes due to decreased stability and reduced cellular protein abundance. These quasi-null variants are major contributors to disease [14–17].

In contrast, edgetic variants do not cause destabilization but affect specific interfaces, abolishing only a subset of interactions, with several examples discussed below. This illustrates that different genetic changes in a protein can lead to similar phenotypic effects through distinct mechanisms, such as destabilization and degradation or specific loss-of-interaction. Conversely, different genetic changes may also lead to distinct phenotypic effects due to specific edgetic perturbations (**Fig. 1**). In the early days of interactomics, proteins that bind many partners at once were termed "party hubs", while those that bind many partners at different points in time were termed "date hubs" [18,19]. Overlapping vs. distinct interaction interfaces are one of the aspects that determine which interaction partners are affected by a variant.

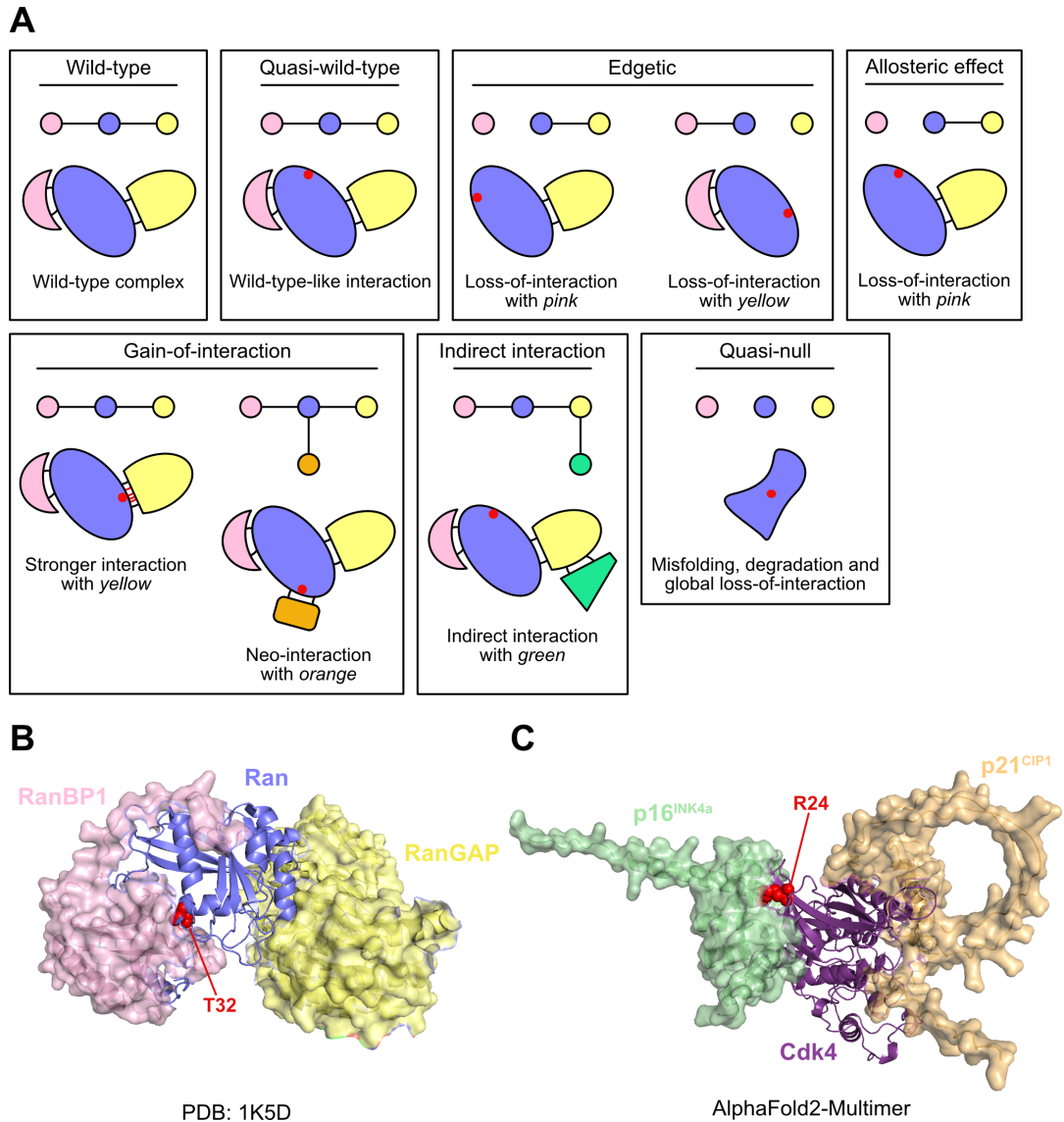

Figure 1. *Schematic illustration of variant effects on protein-protein interactions.* (A) Interaction network illustrations are shown at the top, with the corresponding protein-blob model showing direct and indirect interactions underneath. Amino acid substitutions (red dot) in proteins can have diverse effects on different interaction interfaces, depending on their type and location. The consequences range from no effect (quasi-wild-type) to loss of interaction with specific partners – either through direct perturbations (edgetic) or indirect perturbations (allosteric effect or indirect interaction) – to increased interaction or formation of new interactions (gain-of-interaction), and finally to complete loss of stability and interaction with all partners (quasi-null). (B) In Ran, T32 is located at the interface with RanBP1. Substitution to glutamic

acid disrupts this interaction, but also impairs binding to RanGAP, which binds via a distal interface, indicating allosteric effects of the variant. However, the T32C variant retains binding to RCC1, a third interaction partner of Ran (not shown). This makes T32C an example of an edgetic variant with both direct and allosteric effects on binding [20,21]. (C) Substitution of R24 to cysteine in Cdk4 specifically disrupts interaction with p16$^{INK4a}$, but not with p21$^{CIP1}$, representing a classical edgetic variant [22,23].

While our focus here is on variants affecting protein interactions specifically, there are other crucial aspects to protein function beyond stability and interaction, such as localization, catalytic activity, etc., and losing just one of them can lead to a loss-of-function phenotype. Therefore, a missense variant can affect protein function differently depending on the amino acid substitution and its location within the protein. In many cases, the molecular mechanism underlying disease-causing missense variants can be explained by specific loss of one or more of these properties at the individual protein level [24], which may facilitate development of treatments [16,25,26]. From a diagnostic and variant classification perspective, it is important to consider that pathogenic protein variants may exhibit wild-type-like behaviour with respect to some properties but not others. Hence, not every method will capture all possible consequences on the molecular or cellular level. This review will focus on protein variants that affect protein-protein interactions (PPIs) and highlight the clinical importance of disentangling variant effects on protein function.

**Edgetic perturbations in human diseases**

To elucidate the clinical relevance of disentangling variant effects on protein stability and PPIs, Sahni et al. [27] created libraries consisting of 2,890 disease-associated variants in 1,140 human genes, as well as their wild-type counterparts. They first employed quantitative high-throughput LUMIER assays to test how 2,332 variants impacted protein folding and stability by using the interaction with seven promiscuous chaperones as a proxy for protein destabilization and misfolding. This revealed that around 28% of the variants displayed increased interaction with at least one of the seven chaperones considered in the assay, indicating likely loss of stability. Since more than 70% of the disease-associated variants appeared stable and not misfolded, the authors next used yeast two-hybrid assays to examine how the variants in these proteins affected the interaction with around 7,200 human proteins. This provided detailed interaction profiles for 460 variants in 220 proteins and showed that the disease-associated variants perturbed around 40% of the tested interactions. In conclusion, Sahni et al. found that around 57% of the tested disease-associated variants lead to loss of PPIs, and around 55% of these exerted their effects through specific edgetic perturbations rather than global protein destabilization and misfolding. Additionally, they showed that edgetic variants were significantly enriched in structurally exposed regions and interfaces compared to quasi-null variants, supporting that these variants are likely to disrupt specific interactions. As a control, they observed that around 91% of non-disease-associated variants were quasi-wild-type with respect to both protein stability and PPIs. In another study, Zhong et al. [28] used yeast two-hybrid assays to evaluate the interaction of 29 variants in five proteins implicated in various genetic diseases with 8,076 human proteins. This revealed that 16 variants (55%) caused specific edgetic perturbations. Taken together, these studies emphasize that PPIs are very sensitive to genetic perturbations and explain why a multitude of diseases have been associated with disrupted PPIs. For example, edge-specific mutations in the human cyclin-dependent kinase 4 (CDK4) have been linked to hereditary melanoma. CDK4 phosphorylates and inactivates the retinoblastoma protein and thus promotes $G_1/S$ transition and drives cell-cycle progression and cell proliferation [29]. The activity of CDK4 is inhibited by specific cyclin-dependent kinase inhibitors such as members of the INK4 and CIP/KIP families [30]. Studies have shown that the CDK4 R24C variant specifically disrupts the interaction with $p16^{INK4a}$ and $p18^{INK4c}$, but does not affect the interaction with other family members such as $p15^{INK4b}$ and $p19^{INK4c}$ (INK4 family), and $p21^{CIP1}$ and $p27^{KIP1}$ (CIP/KIP family) (**Fig. 1**) [22,23,28]. In other cases, edge-specific mutations have been associated with distinct phenotypic effects and varying disease severities. For example, edge-specific mutations in the encoding exosome component 3 (EXOSC3) have been associated with varying severity of the rare neurodegenerative disorder Pontocerebellar hypoplasia type 1 (PCH1) [31]. It has been shown that the EXOSC3 G31A variant disrupts the interaction with EXOSC5 but not EXOSC9 and EXOSC1, whereas the EXOSC3 D132A variant disrupts the interaction with EXOSC9 but not EXOSC5 and EXOSC1 [32]. This illustrates that different variants can impact specific interactions while remaining quasi-wild-type for others. Indeed, EXOSC3 G31 and D132 are located in the non-overlapping interfaces with EXOSC5 and EXOSC9, respectively [32–34]. The EXOSC3 G31A variant causes a severe form of PCH1, with patients rarely surviving infancy, while the D132A variant typically allows survival into childhood [35–37]. These examples highlight the importance of detailed mechanistic insights into variant effects on PPIs for diagnosing and characterizing diseases, understanding complex phenotypic outcomes, and optimizing patient treatment.

**Experimental methods to assess protein variant effects on PPIs**

Variant effects on protein function have traditionally been studied in low throughput, focusing on individual variants that have already been clinically observed. Given the vast number of possible and already observed genetic variations in the human genome [38], this one-by-one approach is not feasible for experimentally determining the effects of millions of identified or yet-to-be-seen variants. Multiplexed assays of variant effects (MAVEs), also known as deep mutational scanning (DMS), offer scalable high-throughput methods that can evaluate up to one million variants in a single experiment [39]. Briefly, MAVEs utilize site-saturation mutagenesis libraries and specialized assays to link variant effects on protein function with frequencies determined by next-generation sequencing. This allows the calculation of a score for each variant that reflects its functionality relative to the wild-type. Most MAVEs follow similar workflows, including library generation, transfection or transformation, selection, amplicon generation, deep sequencing, and data analysis.

In recent years, numerous MAVEs on clinically relevant proteins have generated a wealth of newly annotated variant effects which have expanded our knowledge of variant effects and diseases [13,40–51]. However, the power and scope of each MAVE depends largely on the selection assay, which determines the specific property that will be tested. Most MAVEs involve growing cells under selective conditions where protein function is linked to cellular growth rates. Some MAVEs assess protein function broadly and thus cover important clinical aspects, however, these will typically not provide detailed mechanistic insights into variant effects. To gain a mechanistic understanding, MAVEs must use tailored selection assays that isolate the impact of mutations on specific protein functions. Given the critical roles of protein stability and activity, many MAVEs have focused on these properties [13,42,49,52,53]. Yet, only a few studies so far have investigated the mutational effects on PPIs [54–56].

The classical yeast two-hybrid (Y2H) system is a well-established technique originally developed to identify novel binary protein-protein interactions [57]. The technique exploits the multi-domain structure of the Gal4 transcription factor to drive the expression of desired reporter genes only if two proteins of interest interact. Briefly, two proteins of interest (bait and prey) are fused to the DNA-binding domain and activation domain of the Gal4 transcription, respectively. If the bait and prey proteins interact, the activation domain is brought in sufficiently close proximity to the DNA-binding domain to reconstitute the functional Gal4 transcription factor that recruits the RNA polymerase II transcription machinery to initiate transcription of reporter genes such as *HIS3* and *URA3*. This allows only yeast cells that express two interacting proteins to grow under selective conditions. Thus, the read-out is an indirect measurement that links the interaction between bait and prey proteins to cell growth via the expression of reporter genes. If not properly controlled for, it can be prone to produce false negative and false positive PPIs. For example, it is important to consider that the interaction must occur in the yeast nucleus to be detected and that the bait protein itself must not operate as an activator of transcription [58].

Y2H has been proven to be a robust and scalable *in vivo* method capable of identifying thousands of PPIs on a proteome-wide scale across multiple organisms. For instance, Y2H was used to create the human reference interactome map, which contains around 53,000 binary PPIs [59]. Beyond identifying numerous new PPIs and generating interaction networks, Y2H can also be utilized for high-throughput investigation of individual PPIs. In these experiments, Y2H can assess how variants in one protein affect its interaction with a partner. Specifically, yeast cells are co-transformed with a wild-type bait protein and a library of missense variants of the prey protein. This approach allows identifying which positions in the prey protein are sensitive to mutations and likely critical for mediating the interaction with the bait protein. However, it is important to consider that protein variants might lose the interaction with the bait protein for other reasons than directly disrupting critical intermolecular contacts between the bait and prey proteins. For example, protein variants may lead to loss of stability, mislocalization, or interfere with a scaffold protein essential for bridging the interaction between the bait and prey proteins (**Fig. 1**). In Y2H, these different effects cannot be distinguished, as they all present the same phenotype. Therefore, to separate those variants that lose interaction due to loss of stability from those that lose interaction due to actual loss of binding, Y2H assays should be complemented with additional assays tailored to probe protein stability. This approach has been pursued in Faure et al. [60] where they applied combinations of high-throughput protein-fragment complementation assays [61–63], termed abundancePCA and bindingPCA, to evaluate variant effects on protein stability and PPIs, respectively. They refer to this combined approach as doubledeepPCA and applied it to two PPI domains: the C-terminal SH3 domain of GRB2 (interacting with GAB2) and the third PDZ domain of PSD-95 (interacting with CRIPT). This method can ideally identify quasi-wild-type, edgetic and quasi-null variants for the studied PPIs. In a recent study, we applied a similar approach to study the effects of MLH1 variants on protein stability and interaction with its partner, PMS2 [56]. MLH1 is involved in DNA mismatch repair, and loss-of-function variants in the MLH1 protein are known to cause Lynch syndrome, a disease that increases the risk of several cancers, particularly colorectal cancer [64–66]. We found that MLH1 variants often show loss of interaction with PMS2 due to loss of stability (example of quasi-null variants). Interestingly, some variants in the central region of MLH1, far from the known PMS2-binding interface, led to markedly increased interaction (example of gain-of-interaction variants). However, it remains to be determined whether this reflects a novel binding interface, possibly involving the disordered linker region in PMS2, or an allosteric effect propagated through the structure that enhances interaction at the known PMS2-binding interface.

Both classical Y2H assays and doubledeepPCA typically focus on probing the mutational effects for individual PPIs. However, proteins may have multiple interaction partners and variants can have varying effects on these interactions. In a recent study, Bendel et al. [67] used bindingPCA to measure how variants in the bZIP domain protein JUN affected its interaction with all other members of the conserved bZIP superfamily. In brief, they systematically assessed all possible missense variants in 32 positions of JUN for interactions with 52 other bZIP proteins, resulting in 26,648 interaction pairs. Their findings provide insights into binding affinity and specificity and showed that many JUN variants disrupted binding with multiple interaction partners. However, certain variants displayed edgetic effects and altered the binding specificity of JUN, leading to either increased or decreased binding with specific partners. Importantly, these variants often also altered the binding affinity of non-affected interactions, which indicates that, at least in small PPI domains, edgetic variants cannot always be defined as quasi-wild-type for other interactions. The approach pursued by Bendel et al. enables parallel characterization of variant effects on interactions with multiple partners. Limitations are that it relies on prior knowledge of interaction partners and is limited to binary interactions.

Affinity purification-mass spectrometry (AP-MS) is an alternative high-throughput method to investigate PPIs [68]. In AP-MS, a tagged bait protein is purified along with its interaction partner(s) from a cellular lysate. The co-purified prey protein(s) are enzymatically digested into peptides, which are analysed by mass spectrometry for identification. This method is not limited to binary interactions and provides an unbiased detection of PPIs and protein complexes, so long the interactions are strong enough for co-purification. AP-MS has been employed to globally map interaction networks in various organisms under native conditions [69–72]. Recently, Rodriguez-Mias et al. [73] utilized AP-MS to develop a technology in baker's yeast aimed at identifying positions in a bait protein where mutations are likely to impair complex formation with other proteins. The approach, termed Miro, uses mistranslation to randomly incorporate non-canonical amino acids with distinct physicochemical properties into cellular proteins, creating so-called statistical proteomes composed of unique protein variants. Rodriguez-Mias and colleagues expressed a tagged version of the ribosomal protein Rpl16B in yeast and used Miro to generate a statistical proteome with a proline analogue, azetidine, followed by affinity purification of Rpl16B under native conditions and digestion of interacting proteins. The resulting peptides were analysed using mass spectrometry to determine the abundance of each peptide containing azetidine relative to its wild-type counterpart, both in the input proteome and after affinity purification. They identified over 1,200 proteins co-purified with Rpl16B, mostly ribosomal proteins. Importantly, the abundance of peptides containing azetidine was significantly reduced after affinity purification, suggesting that azetidine perturbs ribosomal protein assembly. While this approach offers an unbiased method for studying variant effects on protein complex formation, it does not distinguish direct from indirect PPIs and might not capture weak or transient interactions (**Fig. 1**).

**Peptide-mediated interactions**

Many transient interactions are mediated by short linear motifs (SLiMs), which are distinct sequences of 3-10 residues typically found in intrinsically disordered regions (IDRs) of the proteome [74]. Despite their important role in regulating fundamental biological processes, transient, low-affinity SLiM-based interactions are less studied compared to obligate, stable interactions due to technical limitations of conventional methods like Y2H and AP-MS. However, methods tailored to study such interactions have been developed. For example, proteomic peptide-phage display (ProP-PD) has proven a powerful approach to shed light on this unexplored part of the interactome [75,76]. In ProP-PD, a purified bait protein is screened against a library encoding custom proteomic peptides displayed on the surface of phage particles. Importantly, several copies of the same peptide are displayed on a single phage particle, which offers a high avidity and enables detection of transient, low affinity interactions. Phages that display a peptide that binds the bait protein are enriched, and the peptide sequence can be determined using next-generation sequencing. In a recent study, Kliche et al. [77] applied ProP-PD to investigate how disease-causing variants in IDRs affect

SLiM-based interactions. They created a peptide-phage-display library encoding more than 12,000 mutations suspected of being disease-causing. Kliche and colleagues screened 80 diverse bait proteins against this library and identified several mutations that caused either loss-of-interaction or gain-of-interaction. Notably, Kliche et al. showed that a disease-causing variant in CDC45, which caused loss of interaction at the peptide level, had similar effects in the context of the full-length protein, supporting the utility of the peptide-level data and providing insights into the potential disease mechanism for that specific variant. Some variants led to loss of interaction with one bait protein while gaining interaction with another. This demonstrates the approach's capacity to identify mutations that create new binding sites or enhance binding affinity, a category of mutations that so far has received less attention in the field and is more challenging to probe experimentally or computationally, due to the huge number of potential new binding partners (**Fig. 1**) [78,79].

**Computational approaches for predicting the effects of variants on PPIs**

Atom-level resolution models of variants that affect protein-protein interactions (PPIs) can be immensely valuable, particularly in drug development. Databases like dSysMap or Interactome INSIDER [32,80] have gathered experimentally resolved structures of PPIs in human and mapped variants with clinical impact onto them, providing a starting point for exploration of interface consequences and drug development. Yet, despite the dramatic upscaling in experimental throughput, it is unlikely that we will be able to test all interactions of interest experimentally anytime soon, let alone resolve the wild-type and variant structures. Thus, computational approaches that accurately predict the impact of a mutation on a known interaction would be of great practical value. For reasons outlined below, while multiple methods exist, there are several challenges in developing them into generally applicable tools of high accuracy.

Beyond their practical utility, modelling approaches – especially physics-based ones – allow integration of individual data points from experimental measurements with the biophysical models of protein structure, stability and interactions. Such models generate testable hypotheses about how a new variant or mutation will affect a given interaction, thereby enabling learning loops that can improve accuracy in a targeted manner. For example, a well-described model can identify underdetermined parameters, which may then be specifically addressed through experiments, or suggest which binding partner is most likely affected by a variant. Such investigations may also help uncover higher-order rules that are essential for taking modelling to the next abstraction level [81].

Generally, the performance of computational predictors is limited by the quality and quantity of the available data. In the case of changes in binding affinity of PPIs upon mutation, the data is highly complex. Protein interfaces are very diverse, as are the interactions between proteins and the functions they facilitate. The patterns for amino acids in interfaces differ from those of non-interface regions, and are also different from protein cores, which have been extensively studied in the context of folding and mutational effects on it [82,83]. Structural patterns also vary greatly [84]. As recently reviewed, among current methods predicting protein interfaces, using a structure as input leads to higher accuracy than using a sequence, and knowing the interaction partner yields better results than trying to identify all possible interfaces on the protein surface [85]. To date, most methods require a structure of the protein complex as input [86–94], with sequences used mainly to retrieve additional information, usually conservation-related [95–98], although sequence-based methods are becoming more popular [99,100], often building on the foundation of large protein language models [101,102].

Predictors of changes in folding stability upon mutation used to be limited by the need for an experimentally solved structure. This has changed with the release of AlphaFold2 [103], which makes accurate prediction of structure from sequence possible for many folded proteins [104]. It has been shown that AlphaFold2-predicted structures can be used as input for stability predictors without substantial changes in accuracy compared to experimental structures [104–106]. Prediction of protein complex structures has not reached the desired levels of precision yet, with AlphaFold-Multimer [107] and more recently released AlphaFold3 [108] showing only moderate accuracy for multimeric structure predictions [109–111], and self-reporting metrics do not always reflect the accuracy interface prediction [112–114]. Several sequence-based machine learning methods were developed for binding affinity change predictions before AlphaFold's release, such as SAAMBE-SEQ [115] and ProAffiMuSeq [116]. Recently, DeepPPAPredMut [117] makes use of AlphaFold-Multimer for structure-based feature generation, and the ESM group of models is being used as sequence-feature input [101,102]. Interestingly, both ProAffiMuSeq and DeepPPAPredMut achieve improvements in correlation to experimental data by dividing PPIs into functional classes, such as antigen–antibody and enzyme-inhibitor. This might indicate that models perform better when the complex and diverse landscape of PPIs is broken down into specific groups with well-defined properties. For example, narrowly-specialized tools like FlexPepDock [118], developed specifically for generating precise protein-peptide complexes, consider the increased flexibility of peptides and other distinct traits of peptide-protein interactions. Apart from the structural data and auxiliary information, the main component of the training data for the binding affinity change predictors is the experimental data on the binding affinity of complexes before and after mutation. Before the first version of the SKEMPI database [119], which integrates binding affinity changes with PDB structure references and aggregates other datasets [120,121], this type of data was not systematically available. Most binding affinity change predictors developed before that extrapolated from physicochemical energy calculations for changes in monomeric stability upon mutation and were not specifically fitted for binding affinity change estimation. However, binding, compared to folding, has a much smaller energy gap between the optimal native state and alternatives, making it harder to model [122]. Widely used methods of this type are parts of the Rosetta [89,123] and FoldX [86,87] macromolecular modelling suites. Both have physics-inspired energy functions at their core, with various applications utilizing these functions to perform different tasks of protein modelling and design. The energy functions consist of different partial score terms aimed at representing the main intermolecular forces, mainly acting on bond lengths and angles between atoms [86,124,125]. Each partial score has a weight associated with it, determined by fitting to experimental data. Thus, one important characteristic of the algorithm is the data on which it was calibrated [89,124]. The FoldX energy function was fitted to the first versions of the ProTherm database [126], which contains data on mutation-caused stability changes in monomers. The Rosetta energy functions were primarily designed not for PPIs but for other tasks, such as predicting protein folding by distinguishing between optimally and sub-optimally folded structures, generating stable *de novo* designs, and predicting the effect of mutations on protein stability. Several Rosetta applications can estimate changes in binding affinity upon mutation. Flex ddG [88] is the most recent Rosetta method developed specifically for this purpose, scoring the wild-type and mutant protein complexes twice, in bound and unbound states, to calculate the energy of binding. FoldX's AnalyseComplex function uses the same principle. BeAtMuSiC [93] also combines values of stability change for the complex and individual partners to estimate changes in binding affinity.

More recently, BindProfX [97] and MutaBind [96,127] combine energy terms with evolutionary information. BindProfX integrates FoldX interface scores with an evolutionary interface structure profile, while MutaBind uses evolutionary conservation as one of its score terms. In many recent bioinformatics applications, evolutionary conservation has proven to be a good estimator for the effects of mutations on fitness, with many methods utilizing multiple sequence alignment (MSA) information for variant effect prediction [128,129]. Typically, conservation-based methods predict an overall fitness of a variant without providing insights into specific mechanisms. By combining conservation scores with predictions of changes in protein stability, variants with poor conservation scores can be divided into two categories: those predicted to be non-functional and unstable, and those predicted to impair function without affecting stability [130]. Interaction-disrupting variants fall into the second category; however, the local modelling applied in most protocols may make it challenging to pick up those disruptions stemming from allosteric effects. This coupling of two types of predictors has been used to classify mutations by mechanism of effect [17,131] and may prove useful in the more general prediction of variant effects upon mutation.

Machine learning approaches for protein structure prediction are revolutionizing the field [103,132] and are also becoming widely used for predicting stability changes upon mutation [105,133–135]. The release of SKEMPI 2.0 [136] has more than doubled the number of unique entries. However, as noted in publications describing new tools [90,137] and benchmarking works [138,139], the quantity and quality of the data mean that making accurate predictions of binding affinity changes remains a challenging task: more than half of the mutations in SKEMPI 2.0 are to alanine, and this trend persists when the database is filtered to include only single-point mutations (**Fig. 2**). Given 20 protein-coding amino acids, there are 380 possible mutation types. If mutation types were equally distributed in S4169, one of the most used SKEMPI 2.0 single-point subsets [91,92,95], 4169/380 would give us roughly 11 data points per mutation type. In reality, 67 types are not represented at all (**Fig. 2**). Furthermore, a diverse dataset would also include variations in the context surrounding the mutation site, whereas the SKEMPI database self-reports a bias towards certain complexes and interactions, such as protease-inhibitor and antibody-antigen complexes. Yet another obstacle is the measurement error in the data, which inevitably propagates into models. The SKEMPI 2.0 database contains multiple measurements for some mutations where the same variants were tested by multiple groups or techniques. Despite overall agreement between different experiments, the estimated upper bound for the Pearson correlation between predicted and experimental data is 0.89, when calculated based on data variation in SKEMPI 2.0 [139].

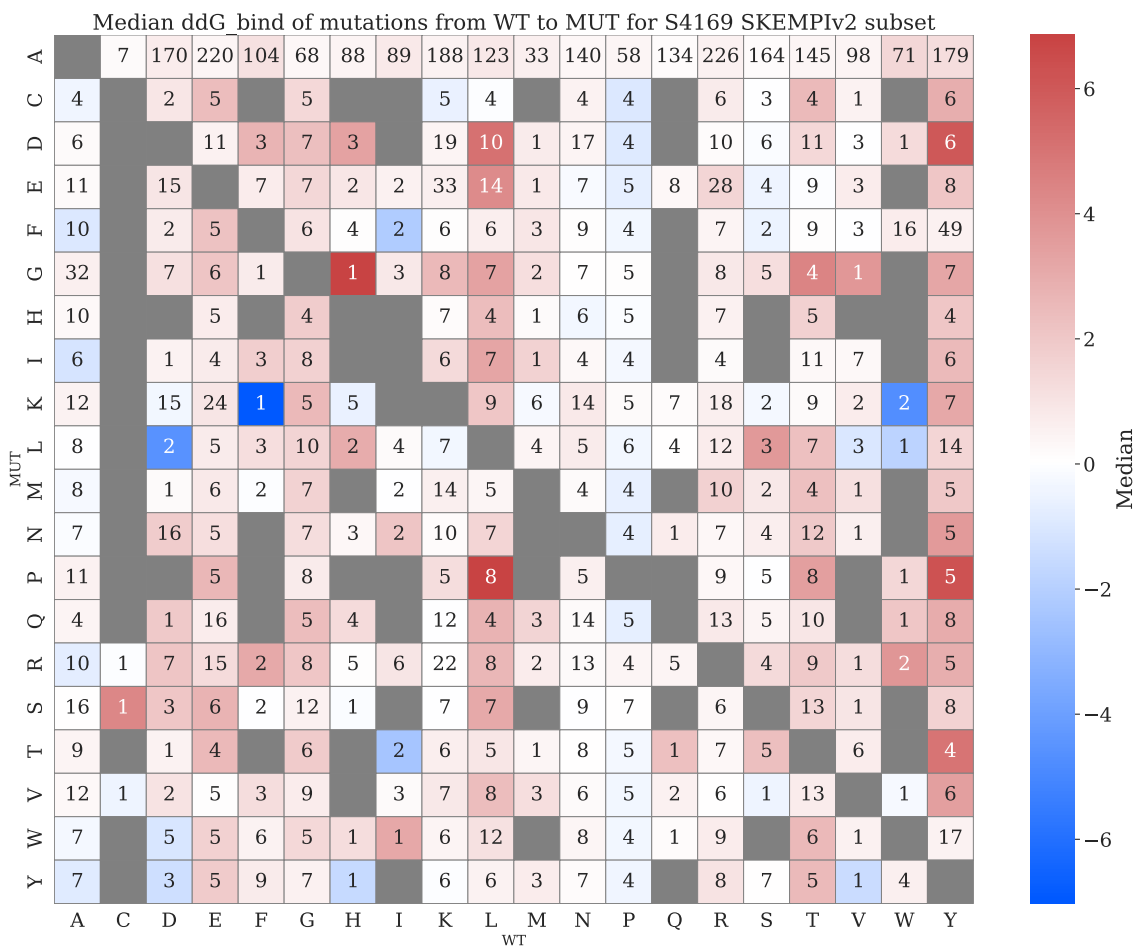

Figure 2. *Counts and median changes in binding affinity for each mutation type in the S4169 subset of the SKEMPI 2.0 database* [136]. The matrix indicates the number of datapoints for each mutation type, with color indicating the median change in stability (measured in kcal/mol, with positive values indicating increased dG of binding upon mutation, and, thus, destabilisation). The x-axis denotes the wild-type residue, while the y-axis denotes the mutant residue. Mutations to alanine are overrepresented, an issue also observed in the ProTherm database [140]; this is due to the historical prevalence of "side chain truncations", in other words, mutations to alanine.

Both in protein binding affinity and stability change datasets (like the ProTherm database [126]), we observed a prevalence of mutations that lead to destabilization of the protein and/or interaction over the ones that lead to stabilization. In the case of folding stability, this can be partially explained by the fact that most proteins have evolved to maintain only a marginally stable conformation while remaining functional (with the difference between unfolded and folded states of most proteins being within the range of 10 kcal/mol), [141,142], thus being easily destabilized even by a single point mutation [143]. In the case of the binding affinity data, interaction strength is fine-tuned for the specific function of the protein, requiring not the strongest, but optimal binding affinity. However, protein complexes gathered by SKEMPI 2.0 have strong enough interactions to be co-purified for experimental structure solving, introducing the bias. In addition to that, mutations to alanine, which are prevalent in the SKEMPI 2.0 database due to the popularity of alanine scanning experiments, rarely improve binding affinity [144] (**Fig. 2**). Method developers attempt to compensate for this bias by augmenting data with reversed mutations [95,96], as changing back from the variant to the wild-type residue should result in the opposite binding affinity change. For example, S4169 is expanded into S8338 by reversing every mutation in the set. This augmentation has, however, not yet led to substantial improvement in overall performance. One reason for this may be that starting structures for the wild-type proteins in SKEMPI 2.0 are available – that is the core requirement for the variants in the database, while starting structures for the reverse mutations typically are not. Predicted models can be used instead, but unless specifically benchmarked [140], it is unclear how the use of models versus experimentally resolved structures affects performance. To improve prediction accuracy by introducing more backbone variability, Samuel Coulbourn Flores, Athanasios Alexiou et al. [145] developed a workflow that runs predictions on homologous structures and averages the results. Mutations, especially those involving substantial changes in side-chain volume, charge, or hydrophobicity, can induce local and sometimes more extensive backbone rearrangements to accommodate the new residue [146–148]. Including backbone flexibility into the model can improve binding affinity change estimations [88,149,150]. Another approach used in recent works is self-supervised learning on the protein complexes before training for binding affinity prediction, enabling neural networks to recognize patterns in local environments at the interfaces for the broader set of examples than what is present in SKEMPI 2.0 [90,91].

For performance assessment, it is also important to note that all the machine learning methods discussed in this review, as well as most of those omitted due to space constraints, are trained and tested on the SKEMPI 2.0 database, which makes independent accuracy assessment very challenging [139]. While cross-validation is an important practice for evaluating performance, it would be necessary to restrict the train-test split to the protein interaction level rather than just the entry level or PDB ID level, as 37 out of 237 PPIs in the SKEMPI 2.0 database are represented by more than one PDB structure. Otherwise, the model may be trained and validated on the same or closely related protein complex structure, leading to inflated correlations with experimental data. This issue affects both machine learning methods and traditional modeling approaches. The SKEMPI 2.0 dataset provides information to address the problem. The authors identify homologous interactions for every entry in the dataset using sequence similarity (details can be found in the original paper [110]). For some entries (~15%), there are no homologous interactions, but most have at least one homolog (**Fig. 3B**). There are three large manually assigned clusters of homologous interactions: protease-inhibitor, pMHC-TCR, and antibody-antigen. In order to avoid overfitting, the SKEMPI 2.0 article suggests keeping homologous interactions out of the train set when testing on a structure from that homologous group. While this additional information is very important for the proper train-test splits, the relatively small size of the dataset means that putting one of the members of the three big clusters into the test set will immediately reduce the training set size by 10 to 20% [111]. Splitting by PDB structure for validation has a similar problem: although more than half of the PDB structures in the dataset have less than 10 entries per structure, random splitting will result in unevenly distributed mutation types in the training set per round of validation (**Fig. 3A**). Structure- and homology-aware splitting thus becomes a double-edged sword, avoiding inflations of the accuracy scores but leading to further imbalances in the training set.

A good baseline for assessing the quality of predictive method performance is comparing it to a random predictor (e. g. in ROC curves), a uniform predictor assigning the same value to all test cases, or another simplified model. For example, for ProTherm, which contains data on the mutational effect on folding stability instead of binding, it has been shown that a Pearson's correlation coefficient as high as 0.42 can be achieved by a model that simply predicts stability change to be the mean of all mutations of the same wild-type and target amino acid [118]. An analogous analysis of S4169 gave a substantially lower correlation coefficient of 0.29 for the mutation-level split, which could be used as baseline accuracy for binding affinity change predictors. The difference between SKEMPI 2.0 and ProTherm, in this case, might be caused by the higher complexity of the binding affinity change data, as described above. Additionally, ProTherm contains a more diverse set of mutations, most notably, mutations to glycine are almost as abundant as those to alanine.

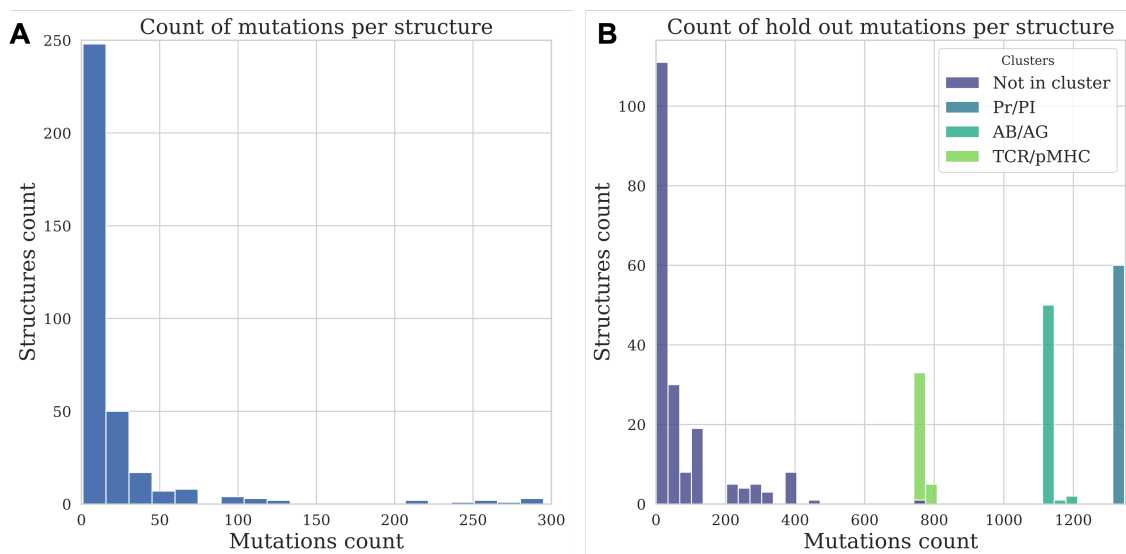

Figure 3. *Mutation counts distribution in the full SKEMPI2 dataset.* Histograms of (A) the mutation (entry) counts per PDB structure and (B) the count of mutations excluded from the train set following SKEMPI2 homologous interaction labelling. (A) All entries are grouped by the PDB structure, and mutations per structure are counted. (B) All entries are grouped by the PDB structure, and for each structure, we count the number of mutations to be excluded from the train set by adding up all entries for this structure and structures labelled as homologous directly or by belonging to the same homology cluster. Two structures are labelled as both Pr/PI and AB/AG and would require excluding 2486 mutations from the training set.

One way to overcome the limitations of currently available databases may be to use data from high-throughput experiments, such as those described above. This approach will require addressing challenges of assay dependencies and standardization, such as the impact of protein concentration on experimental readouts [151]. However, it could provide much more balanced datasets than are currently available (**Fig. 2**). Previous studies focusing on mutational effects in monomers have successfully pooled data from different MAVEs after normalization [130,152]. Methods for translating experimentally determined MAVE scores into binding affinities have been developed, particularly by Heyne et al. [144]. An analogous pooling approach for protein interactions could be promising and may provide a more balanced training dataset for future methods.

**Outlook**

With high-throughput screens producing data at unprecedented volumes and advances in protein modeling, it is very likely that models of PPIs and the consequences of variants on these interactions will improve in the coming years. Ideally, there will be synergy between the broader, less biased databases provided by high-throughput experiments and the advances in capturing the details and key features of protein interactions.

Experimental approaches that disentangle whether a variant affects the interaction directly or if it destabilizes the protein, leading to a quasi-null phenotype, will help fine-tuning prediction accuracy. Further, studies that test the effects of variants on multiple interaction partners may provide a more systematic overview of variant consequences, helping to avoid the limitations of screening specific partners [56,153]. The increasing complexity and volume of modern screens, as well as the complexity of cellular wiring that protein interaction networks represent, underscore the need for greater integration between experimental testing and modeling.

The approaches discussed in this review focus on consequences of variants on direct, physical protein-protein interactions. There are many other aspects that may affect whether two proteins – whether wild-type or variant – will interact in a cell, including expression patterns and localisation.

Variants that lead to gain of function remain a particular challenge. In the form of gain-of-interaction variants, they would lead to newly arising interfaces and interactions. While methods like Y2H assays can easily detect loss of interaction, identifying new interactions is much more difficult unless the study specifically targets suspected new partners. AP-MS-based methods have a better chance of detecting new partners, although they can be affected by noise. Both methods operate within dynamic ranges determined by the specific experimental setup that may affect whether low-affinity binding or increases in affinity are accurately detected. Similarly, computational methods can capture detrimental effects on a known interface relatively easily. However, *in silico* screening for novel interactions is computationally expensive [154], and the reliability of these methods often depends on the sensitivity and specificity of the underlying approach. Because established cases of gain-of-interaction variants are relatively scarce, the effectiveness of these methods is often not well understood.

Going beyond proteins, screens that probe the effects of small molecules on protein interactions will allow insights into the complex interplay of different biomolecules, a promising area of drug development [155]. Integrative models that bridge biology and chemistry will be crucial for advancing drug design.

**Acknowledgments**

We thank our colleagues at the Linderstrøm-Lang Centre for Protein Science for insightful discussions and the Lundbeck Foundation and Novo Nordisk Foundation for financial support.

**Declaration of interests**

The authors declare no competing interests.

**References**

1. Apgar TL, Sanders CR. Compendium of causative genes and their encoded proteins for common monogenic disorders. Protein Science. 2022 Jan 21;31(1):75–91.

2. Weile J, Sun S, Cote AG, Knapp J, Verby M, Mellor JC, et al. A framework for exhaustively mapping functional missense variants. Mol Syst Biol. 2017 Dec 21;13(12).

3. Landrum MJ, Lee JM, Benson M, Brown GR, Chao C, Chitipiralla S, et al. ClinVar: improving access to variant interpretations and supporting evidence. Nucleic Acids Res. 2018 Jan 4;46(D1):D1062–7.

4. Chen E, Facio FM, Aradhya KW, Rojahn S, Hatchell KE, Aguilar S, et al. Rates and Classification of Variants of Uncertain Significance in Hereditary Disease Genetic Testing. JAMA Netw Open. 2023 Oct 25;6(10):e2339571.

5. Hockings JK, Pasternak AL, Erwin AL, Mason NT, Eng C, Hicks JK. Pharmacogenomics: An evolving clinical tool for precision medicine. Cleve Clin J Med. 2020 Feb;87(2):91–9.

6. Cusick ME, Klitgord N, Vidal M, Hill DE. Interactome: gateway into systems biology. Hum Mol Genet. 2005 Oct 15;14(suppl_2):R171–81.

7. Vidal M, Cusick ME, Barabási AL. Interactome Networks and Human Disease. Cell. 2011 Mar;144(6):986–98.

8. Rual JF, Venkatesan K, Hao T, Hirozane-Kishikawa T, Dricot A, Li N, et al. Towards a proteome-scale map of the human protein–protein interaction network. Nature. 2005 Oct 28;437(7062):1173–8.

9. Sahni N, Yi S, Zhong Q, Jailkhani N, Charloteaux B, Cusick ME, et al. Edgotype: a fundamental link between genotype and phenotype. Curr Opin Genet Dev. 2013 Dec;23(6):649–57.

10. Hingorani KS, Gierasch LM. Comparing protein folding in vitro and in vivo: foldability meets the fitness challenge. Curr Opin Struct Biol. 2014 Feb;24:81–90.

11. Grønbæk-Thygesen M, Voutsinos V, Johansson KE, Schulze TK, Cagiada M, Pedersen L, et al. Deep mutational scanning reveals a correlation between degradation and toxicity of thousands of aspartoacylase variants. Nat Commun. 2024 May 13;15(1):4026.

12. Clausen L, Voutsinos V, Cagiada M, Johansson KE, Grønbæk-Thygesen M, Nariya S, et al. A mutational atlas for Parkin proteostasis. Nat Commun. 2024 Feb 20;15(1):1541.

13. Matreyek KA, Starita LM, Stephany JJ, Martin B, Chiasson MA, Gray VE, et al. Multiplex assessment of protein variant abundance by massively parallel sequencing. Nat Genet. 2018 Jun 21;50(6):874–82.

14. Casadio R, Vassura M, Tiwari S, Fariselli P, Luigi Martelli P. Correlating disease-related mutations to their effect on protein stability: A large-scale analysis of the human proteome. Hum Mutat. 2011 Oct;32(10):1161–70.

15. Taipale M. Disruption of protein function by pathogenic mutations: common and uncommon mechanisms. Biochemistry and Cell Biology. 2019 Feb;97(1):46–57.

16. Stein A, Fowler DM, Hartmann-Petersen R, Lindorff-Larsen K. Biophysical and Mechanistic Models for Disease-Causing Protein Variants. Trends Biochem Sci. 2019 Jul;44(7):575–88.

17. Cagiada M, Johansson KE, Valanciute A, Nielsen S V, Hartmann-Petersen R, Yang JJ, et al. Understanding the Origins of Loss of Protein Function by Analyzing the Effects of Thousands of Variants on Activity and Abundance. Mol Biol Evol. 2021 Jul 29;38(8):3235–46.

18. Agarwal S, Deane CM, Porter MA, Jones NS. Revisiting Date and Party Hubs: Novel Approaches to Role Assignment in Protein Interaction Networks. PLoS Comput Biol. 2010 Jun 17;6(6):e1000817.

19. Chang X, Xu T, Li Y, Wang K. Dynamic modular architecture of protein-protein interaction networks beyond the dichotomy of 'date' and 'party' hubs. Sci Rep. 2013 Apr 22;3(1):1691.

20. Perica T, Mathy CJP, Xu J, Jang GM, Zhang Y, Kaake R, et al. Systems-level effects of allosteric perturbations to a model molecular switch. Nature. 2021 Nov 4;599(7883):152–7.

21. Seewald MJ, Körner C, Wittinghofer A, Vetter IR. RanGAP mediates GTP hydrolysis without an arginine finger. Nature. 2002 Feb;415(6872):662–6.

22. Wölfel T, Hauer M, Schneider J, Serrano M, Wölfel C, Klehmann-Hieb E, et al. A p16INK4a-insensitive CDK4 mutant targeted by cytolytic T lymphocytes in a human melanoma. Science. 1995 Sep 1;269(5228):1281–4.

23. Rane SG, Dubus P, Mettus R V., Galbreath EJ, Boden G, Reddy EP, et al. Loss of Cdk4 expression causes insulin-deficient diabetes and Cdk4 activation results in β-islet cell hyperplasia. Nat Genet. 1999 May;22(1):44–52.

24. Taipale M. Disruption of protein function by pathogenic mutations: common and uncommon mechanisms. Biochemistry and Cell Biology. 2019 Feb;97(1):46–57.

25. Nada H, Choi Y, Kim S, Jeong KS, Meanwell NA, Lee K. New insights into protein–protein interaction modulators in drug discovery and therapeutic advance. Signal Transduct Target Ther. 2024 Dec 6;9(1):341.

26. Zhu H, Gao H, Ji Y, Zhou Q, Du Z, Tian L, et al. Targeting p53–MDM2 interaction by small-molecule inhibitors: learning from MDM2 inhibitors in clinical trials. J Hematol Oncol. 2022 Dec 13;15(1):91.

27. Sahni N, Yi S, Taipale M, Fuxman Bass JI, Coulombe-Huntington J, Yang F, et al. Widespread Macromolecular Interaction Perturbations in Human Genetic Disorders. Cell. 2015 Apr;161(3):647–60.

28. Zhong Q, Simonis N, Li Q, Charloteaux B, Heuze F, Klitgord N, et al. Edgetic perturbation models of human inherited disorders. Mol Syst Biol. 2009 Jan 3;5(1).

29. Baker SJ, Poulikakos PI, Irie HY, Parekh S, Reddy EP. CDK4: a master regulator of the cell cycle and its role in cancer. Genes Cancer. 2022 Aug 25;13:21–45.

30. Sherr CJ, Roberts JM. CDK inhibitors: positive and negative regulators of G1-phase progression. Genes Dev. 1999 Jun 15;13(12):1501–12.

31. Wan J, Yourshaw M, Mamsa H, Rudnik-Schöneborn S, Menezes MP, Hong JE, et al. Mutations in the RNA exosome component gene EXOSC3 cause pontocerebellar hypoplasia and spinal motor neuron degeneration. Nat Genet. 2012 Jun 29;44(6):704–8.

32. Mosca R, Tenorio-Laranga J, Olivella R, Alcalde V, Céol A, Soler-López M, et al. dSysMap: exploring the edgetic role of disease mutations. Nature Methods 2015 12:3 [Internet]. 2015 Feb 26 [cited 2024 Aug 5];12(3):167–8. Available from: https://www.nature.com/articles/nmeth.3289

33. Morton DJ, Kuiper EG, Jones SK, Leung SW, Corbett AH, Fasken MB. The RNA exosome and RNA exosome-linked disease. RNA. 2018 Feb;24(2):127–42.

34. Fasken MB, Losh JS, Leung SW, Brutus S, Avin B, Vaught JC, et al. Insight into the RNA Exosome Complex Through Modeling Pontocerebellar Hypoplasia Type 1b Disease Mutations in Yeast. Genetics. 2017 Jan 1;205(1):221–37.

35. Eggens VR, Barth PG, Niermeijer JMF, Berg JN, Darin N, Dixit A, et al. EXOSC3 mutations in pontocerebellar hypoplasia type 1: novel mutations and genotype-phenotype correlations. Orphanet J Rare Dis. 2014 Dec 13;9(1):23.

36. Schwabova J, Brozkova DS, Petrak B, Mojzisova M, Pavlickova K, Haberlova J, et al. Homozygous *EXOSC3* Mutation c.92G→C, p.G31A is a Founder Mutation Causing Severe Pontocerebellar Hypoplasia Type 1 Among the Czech Roma. J Neurogenet. 2013 Dec 25;27(4):163–9.

37. Biancheri R, Cassandrini D, Pinto F, Trovato R, Di Rocco M, Mirabelli-Badenier M, et al. EXOSC3 mutations in isolated cerebellar hypoplasia and spinal anterior horn involvement. J Neurol. 2013 Jul 7;260(7):1866–70.

38. Karczewski KJ, Francioli LC, Tiao G, Cummings BB, Alföldi J, Wang Q, et al. The mutational constraint spectrum quantified from variation in 141,456 humans. Nature. 2020 May 28;581(7809):434–43.

39. Fowler DM, Fields S. Deep mutational scanning: a new style of protein science. Nat Methods. 2014 Aug 30;11(8):801–7.

40. Findlay GM, Daza RM, Martin B, Zhang MD, Leith AP, Gasperini M, et al. Accurate classification of BRCA1 variants with saturation genome editing. Nature. 2018 Oct 12;562(7726):217–22.

41. Jia X, Burugula BB, Chen V, Lemons RM, Jayakody S, Maksutova M, et al. Massively parallel functional testing of MSH2 missense variants conferring Lynch syndrome risk. The American Journal of Human Genetics. 2021 Jan;108(1):163–75.

42. Mighell TL, Evans-Dutson S, O'Roak BJ. A Saturation Mutagenesis Approach to Understanding PTEN Lipid Phosphatase Activity and Genotype-Phenotype Relationships. The American Journal of Human Genetics. 2018 May;102(5):943–55.

43. Kotler E, Shani O, Goldfeld G, Lotan-Pompan M, Tarcic O, Gershoni A, et al. A Systematic p53 Mutation Library Links Differential Functional Impact to Cancer Mutation Pattern and Evolutionary Conservation. Mol Cell. 2018 Jul;71(1):178-190.e8.

44. Chiasson MA, Rollins NJ, Stephany JJ, Sitko KA, Matreyek KA, Verby M, et al. Multiplexed measurement of variant abundance and activity reveals VKOR topology, active site and human variant impact. Elife. 2020 Sep 1;9.

45. Sun S, Weile J, Verby M, Wu Y, Wang Y, Cote AG, et al. A proactive genotype-to-patient-phenotype map for cystathionine beta-synthase. Genome Med. 2020 Dec 30;12(1):13.

46. Amorosi CJ, Chiasson MA, McDonald MG, Wong LH, Sitko KA, Boyle G, et al. Massively parallel characterization of CYP2C9 variant enzyme activity and abundance. The American Journal of Human Genetics. 2021 Sep;108(9):1735–51.

47. Newberry RW, Leong JT, Chow ED, Kampmann M, DeGrado WF. Deep mutational scanning reveals the structural basis for α-synuclein activity. Nat Chem Biol. 2020 Jun 9;16(6):653–9.

48. Roychowdhury H, Romero PA. Microfluidic deep mutational scanning of the human executioner caspases reveals differences in structure and regulation. Cell Death Discov. 2022 Jan 10;8(1):7.

49. Gersing S, Cagiada M, Gebbia M, Gjesing AP, Coté AG, Seesankar G, et al. A comprehensive map of human glucokinase variant activity. Genome Biol. 2023 Apr 26;24(1):97.

50. Sharo AG, Zou Y, Adhikari AN, Brenner SE. ClinVar and HGMD genomic variant classification accuracy has improved over time, as measured by implied disease burden. Genome Med. 2023 Jul 13;15(1):51.

51. Manolio TA, Fowler DM, Starita LM, Haendel MA, MacArthur DG, Biesecker LG, et al. Bedside Back to Bench: Building Bridges between Basic and Clinical Genomic Research. Cell. 2017 Mar;169(1):6–12.

52. Cagiada M, Johansson KE, Valanciute A, Nielsen S V, Hartmann-Petersen R, Yang JJ, et al. Understanding the Origins of Loss of Protein Function by Analyzing the Effects of Thousands of Variants on Activity and Abundance. Mol Biol Evol. 2021 Jul 29;38(8):3235–46.

53. Suiter CC, Moriyama T, Matreyek KA, Yang W, Scaletti ER, Nishii R, et al. Massively parallel variant characterization identifies *NUDT15* alleles associated with thiopurine toxicity. Proceedings of the National Academy of Sciences. 2020 Mar 10;117(10):5394–401.

54. Weng C, Faure AJ, Escobedo A, Lehner B. The energetic and allosteric landscape for KRAS inhibition. Nature. 2024 Feb 15;626(7999):643–52.

55. Faure AJ, Domingo J, Schmiedel JM, Hidalgo-Carcedo C, Diss G, Lehner B. Mapping the energetic and allosteric landscapes of protein binding domains. Nature. 2022 Apr 7;604(7904):175–83.

56. Larsen-Ledet S, Stein A. Disentangling the mutational effects on protein stability and interaction of human MLH1. bioRxiv [Internet]. 2024 Jul 29 [cited 2024 Aug 1];2024.07.28.605491. Available from: https://www.biorxiv.org/content/10.1101/2024.07.28.605491v1

57. Fields S, Song O kyu. A novel genetic system to detect protein–protein interactions. Nature. 1989 Jul;340(6230):245–6.

58. Shivhare D, Musialak-Lange M, Julca I, Gluza P, Mutwil M. Removing auto-activators from yeast-two-hybrid assays by conditional negative selection. Sci Rep. 2021 Mar 9;11(1):5477.

59. Luck K, Kim DK, Lambourne L, Spirohn K, Begg BE, Bian W, et al. A reference map of the human binary protein interactome. Nature. 2020 Apr 16;580(7803):402–8.

60. Faure AJ, Domingo J, Schmiedel JM, Hidalgo-Carcedo C, Diss G, Lehner B. Mapping the energetic and allosteric landscapes of protein binding domains. Nature. 2022 Apr 7;604(7904):175–83.

61. Remy I, Michnick SW. Clonal selection and *in vivo* quantitation of protein interactions with protein-fragment complementation assays. Proceedings of the National Academy of Sciences. 1999 May 11;96(10):5394–9.

62. Tarassov K, Messier V, Landry CR, Radinovic S, Molina MMS, Shames I, et al. An in Vivo Map of the Yeast Protein Interactome. Science (1979). 2008 Jun 13;320(5882):1465–70.

63. Levy ED, Kowarzyk J, Michnick SW. High-Resolution Mapping of Protein Concentration Reveals Principles of Proteome Architecture and Adaptation. Cell Rep. 2014 May;7(4):1333–40.

64. Jun S, Kim TG, Ban C. DNA mismatch repair system. FEBS J. 2006 Apr 31;273(8):1609–19.

65. Bronner CE, Baker SM, Morrison PT, Warren G, Smith LG, Lescoe MK, et al. Mutation in the DNA mismatch repair gene homologue hMLH 1 is associated with hereditary non-polyposis colon cancer. Nature. 1994 Mar;368(6468):258–61.

66. Momma T, Gonda K, Akama Y, Endo E, Ujiie D, Fujita S, et al. MLH1 germline mutation associated with Lynch syndrome in a family followed for more than 45 years. BMC Med Genet. 2019 Dec 2;20(1):67.

67. Bendel AM, Faure AJ, Klein D, Shimada K, Lyautey R, Schiffelholz N, et al. The genetic architecture of protein interaction affinity and specificity. Nat Commun. 2024 Oct 14;15(1):8868.

68. Morris JH, Knudsen GM, Verschueren E, Johnson JR, Cimermancic P, Greninger AL, et al. Affinity purification–mass spectrometry and network analysis to understand protein-protein interactions. Nat Protoc. 2014 Nov 2;9(11):2539–54.

69. Butland G, Peregrín-Alvarez JM, Li J, Yang W, Yang X, Canadien V, et al. Interaction network containing conserved and essential protein complexes in Escherichia coli. Nature. 2005 Feb;433(7025):531–7.

70. Gavin AC, Aloy P, Grandi P, Krause R, Boesche M, Marzioch M, et al. Proteome survey reveals modularity of the yeast cell machinery. Nature. 2006 Mar 22;440(7084):631–6.

71. Krogan NJ, Cagney G, Yu H, Zhong G, Guo X, Ignatchenko A, et al. Global landscape of protein complexes in the yeast Saccharomyces cerevisiae. Nature. 2006 Mar 22;440(7084):637–43.

72. Ewing RM, Chu P, Elisma F, Li H, Taylor P, Climie S, et al. Large-scale mapping of human protein–protein interactions by mass spectrometry. Mol Syst Biol. 2007 Jan 13;3(1).

73. Rodriguez-Mias et al. Proteome-wide identification of amino acid substitutions deleterious for protein function. bioRxiv. 2022 Apr 9;

74. Davey NE, Van Roey K, Weatheritt RJ, Toedt G, Uyar B, Altenberg B, et al. Attributes of short linear motifs. Mol BioSyst. 2012;8(1):268–81.

75. Ivarsson Y, Arnold R, McLaughlin M, Nim S, Joshi R, Ray D, et al. Large-scale interaction profiling of PDZ domains through proteomic peptide-phage display using human and viral phage peptidomes. Proceedings of the National Academy of Sciences. 2014 Feb 18;111(7):2542–7.

76. Davey NE, Seo M, Yadav VK, Jeon J, Nim S, Krystkowiak I, et al. Discovery of short linear motif-mediated interactions through phage display of intrinsically disordered regions of the human proteome. FEBS J. 2017 Feb 18;284(3):485–98.

77. Kliche J, Simonetti L, Krystkowiak I, Kuss H, Diallo M, Rask E, et al. Proteome-scale characterisation of motif-based interactome rewiring by disease mutations. Mol Syst Biol. 2024 Jul 15;20(9):1025–48.

78. Meyer K, Kirchner M, Uyar B, Cheng JY, Russo G, Hernandez-Miranda LR, et al. Mutations in Disordered Regions Can Cause Disease by Creating Dileucine Motifs. Cell. 2018 Sep;175(1):239-253.e17.

79. Gerasimavicius L, Livesey BJ, Marsh JA. Loss-of-function, gain-of-function and dominant-negative mutations have profoundly different effects on protein structure. Nat Commun. 2022 Jul 6;13(1):3895.

80. Meyer MJ, Beltrán JF, Liang S, Fragoza R, Rumack A, Liang J, et al. Interactome INSIDER: a structural interactome browser for genomic studies. Nature Methods 2018 15:2 [Internet]. 2018 Jan 1 [cited 2024 Aug 5];15(2):107–14. Available from: https://www.nature.com/articles/nmeth.4540

81. Schmit J, Dill KA. Next-Gen Biophysics: Look to the Forest, beyond the Trees. Annu Rev Biophys. 2023 May 9;52(Volume 52, 2023):V–VIII.

82. Schreiber G. CHAPTER 1: Protein-Protein Interaction Interfaces and their Functional Implications. RSC Drug Discovery Series. 2021;2021-January(78):1–24.

83. Yan C, Wu F, Jernigan RL, Dobbs D, Honavar V. Characterization of protein-protein interfaces. Protein Journal [Internet]. 2008 Jan 13 [cited 2024 Aug 3];27(1):59–70. Available from: https://link.springer.com/article/10.1007/s10930-007-9108-x

84. Stites WE. Protein−Protein Interactions: Interface Structure, Binding Thermodynamics, and Mutational Analysis. Chem Rev [Internet]. 1997 [cited 2024 Aug 1];97(5):1233–50. Available from: https://pubs.acs.org/doi/full/10.1021/cr960387h

85. Rogers JR, Nikolényi G, Alquraishi M. Growing ecosystem of deep learning methods for modeling protein–protein interactions. Protein Engineering, Design and Selection [Internet]. 2023 Jan 21 [cited 2024 Aug 2];36:1–19. Available from: https://dx.doi.org/10.1093/protein/gzad023

86. Guerois R, Nielsen JE, Serrano L. Predicting Changes in the Stability of Proteins and Protein Complexes: A Study of More Than 1000 Mutations. J Mol Biol [Internet]. 2002 Jul 5 [cited 2022 Aug 31];320(2):369–87. Available from: https://linkinghub.elsevier.com/retrieve/pii/S0022283602004424

87. Schymkowitz J, Borg J, Stricher F, Nys R, Rousseau F, Serrano L. The FoldX web server: an online force field. Nucleic Acids Res [Internet]. 2005 Jul 1 [cited 2024 Jul 30];33(suppl_2):W382–8. Available from: https://dx.doi.org/10.1093/nar/gki387

88. Barlow KA, Ó Conchúir S, Thompson S, Suresh P, Lucas JE, Heinonen M, et al. Flex ddG: Rosetta Ensemble-Based Estimation of Changes in Protein-Protein Binding Affinity upon Mutation. Journal of Physical Chemistry B [Internet]. 2018 May 31 [cited 2024 Jul 30];122(21):5389–99. Available from: https://pubs.acs.org/doi/full/10.1021/acs.jpcb.7b11367

89. Park H, Bradley P, Greisen P, Liu Y, Mulligan VK, Kim DE, et al. Simultaneous optimization of biomolecular energy function on features from small molecules and macromolecules. J Chem Theory Comput [Internet]. 2016 Dec 12 [cited 2022 Aug 31];12(12):6201. Available from: /pmc/articles/PMC5515585/

90. Mohseni Behbahani Y, Laine E, Carbone A. Deep Local Analysis deconstructs protein–protein interfaces and accurately estimates binding affinity changes upon mutation. Bioinformatics [Internet]. 2023 Jun 30 [cited 2024 Jul 24];39(Supplement_1):i544–52. Available from: https://dx.doi.org/10.1093/bioinformatics/btad231

91. Liu X, Luo Y, Li P, Song S, Peng J. Deep geometric representations for modeling effects of mutations on protein-protein binding affinity. PLoS Comput Biol [Internet]. 2021 Aug 1 [cited 2024 Jul 25];17(8):e1009284. Available from: https://journals.plos.org/ploscompbiol/article?id=10.1371/journal.pcbi.1009284

92. Wang M, Cang Z, Wei GW. A topology-based network tree for the prediction of protein–protein binding affinity changes following mutation. Nature Machine Intelligence 2020 2:2 [Internet]. 2020 Feb 14 [cited 2024 Jul 25];2(2):116–23. Available from: https://www.nature.com/articles/s42256-020-0149-6

93. Dehouck Y, Kwasigroch JM, Rooman M, Gilis D. BeAtMuSiC: prediction of changes in protein–protein binding affinity on mutations. Nucleic Acids Res [Internet]. 2013 [cited 2022 Aug 31];41(Web Server issue):W333. Available from: /pmc/articles/PMC3692068/

94. Geng C, Vangone A, Folkers GE, Li |, Xue C, Alexandre |, et al. iSEE: Interface structure, evolution, and energy-based machine learning predictor of binding affinity changes upon mutations. Proteins: Structure, Function, and Bioinformatics [Internet]. 2019 Feb 1 [cited 2025 Jun 17];87(2):110–9. Available from: https://onlinelibrary.wiley.com/doi/full/10.1002/prot.25630

95. Rodrigues CHM, Myung Y, Pires DEV, Ascher DB. mCSM-PPI2: predicting the effects of mutations on protein–protein interactions. Nucleic Acids Res [Internet]. 2019 Jul 2 [cited 2022 Sep 1];47(W1):W338–44. Available from: https://academic.oup.com/nar/article/47/W1/W338/5494729

96. Zhang N, Chen Y, Lu H, Zhao F, Alvarez RV, Goncearenco A, et al. MutaBind2: Predicting the Impacts of Single and Multiple Mutations on Protein-Protein Interactions. iScience. 2020 Mar 27;23(3):100939.

97. Xiong P, Zhang C, Zheng W, Zhang Y. BindProfX: Assessing Mutation-Induced Binding Affinity Change by Protein Interface Profiles with Pseudo-Counts. J Mol Biol [Internet]. 2017 Feb 3 [cited 2024 Jul 30];429(3):426–34. Available from: https://pubmed.ncbi.nlm.nih.gov/27899282/

98. Huang X, Zheng W, Pearce R, Zhang Y, Zhang Y. SSIPe: accurately estimating protein–protein binding affinity change upon mutations using evolutionary profiles in combination with an optimized physical energy function. Bioinformatics [Internet]. 2020 Apr 15 [cited 2025 Jun 17];36(8):2429–37. Available from: https://dx.doi.org/10.1093/bioinformatics/btz926

99. Jin R, Ye Q, Wang J, Cao Z, Jiang D, Wang T, et al. AttABseq: an attention-based deep learning prediction method for antigen–antibody binding affinity changes based on protein sequences. Brief Bioinform [Internet]. 2024 May 23 [cited 2025 Jun 17];25(4):304. Available from: https://dx.doi.org/10.1093/bib/bbae304

100. Sun X, Wu Z, Su J, Li C. A deep attention model for wide-genome protein-peptide binding affinity prediction at a sequence level. Int J Biol Macromol. 2024 Sep 1;276:133811.

101. Zheng F, Jiang X, Wen Y, Yang Y, Li M. Systematic investigation of machine learning on limited data: A study on predicting protein-protein binding strength. Comput Struct Biotechnol J. 2024 Dec 1;23:460–72.

102. Xie J, Zhang Y, Wang Z, Jin X, Lu X, Ge S, et al. PPI-Graphomer: enhanced protein-protein affinity prediction using pretrained and graph transformer models. BMC Bioinformatics [Internet]. 2025 Dec 1 [cited 2025 Jun 18];26(1):1–15. Available from: https://link.springer.com/articles/10.1186/s12859-025-06123-2

103. Jumper J, Evans R, Pritzel A, Green T, Figurnov M, Ronneberger O, et al. Highly accurate protein structure prediction with AlphaFold. Nature 2021 596:7873 [Internet]. 2021 Jul 15 [cited 2024 Jul 25];596(7873):583–9. Available from: https://www.nature.com/articles/s41586-021-03819-2

104. Akdel M, Pires DEV, Pardo EP, Jänes J, Zalevsky AO, Mészáros B, et al. A structural biology community assessment of AlphaFold2 applications. Nature Structural & Molecular Biology 2022 29:11 [Internet]. 2022 Nov 7 [cited 2023 Jul 11];29(11):1056–67. Available from: https://www.nature.com/articles/s41594-022-00849-w

105. Blaabjerg LM, Kassem MM, Good LL, Jonsson N, Cagiada M, Johansson KE, et al. Rapid protein stability prediction using deep learning representations. Elife. 2023 May 15;12.

106. Pak MA, Ivankov DN. Best templates outperform homology models in predicting the impact of mutations on protein stability. Cowen L, editor. Bioinformatics [Internet]. 2022 Jul 27 [cited 2022 Aug 31]; Available from: https://academic.oup.com/bioinformatics/advance-article/doi/10.1093/bioinformatics/btac515/6650690

107. Evans R, O'Neill M, Pritzel A, Antropova N, Senior A, Green T, et al. Protein complex prediction with AlphaFold-Multimer. bioRxiv [Internet]. 2021 Oct 4 [cited 2021 Dec 6];2021.10.04.463034. Available from: https://www.biorxiv.org/content/10.1101/2021.10.04.463034v1

108. Abramson J, Adler J, Dunger J, Evans R, Green T, Pritzel A, et al. Accurate structure prediction of biomolecular interactions with AlphaFold 3. Nature 2024 630:8016 [Internet]. 2024 May 8 [cited 2025 May 20];630(8016):493–500. Available from: https://www.nature.com/articles/s41586-024-07487-w

109. Yin R, Feng BY, Varshney A, Pierce BG. Benchmarking AlphaFold for protein complex modeling reveals accuracy determinants. Protein Science [Internet]. 2022 Aug 1 [cited 2024 Aug 2];31(8):e4379. Available from: https://onlinelibrary.wiley.com/doi/full/10.1002/pro.4379

110. Zhu W, Shenoy A, Kundrotas P, Elofsson A. Evaluation of AlphaFold-Multimer prediction on multi-chain protein complexes. Bioinformatics [Internet]. 2023 Jul 1 [cited 2024 Aug 2];39(7). Available from: https://dx.doi.org/10.1093/bioinformatics/btad424

111. Kovtun D, Akdel M, Goncearenco A, Zhou G, Holt G, Baugher D, et al. PINDER: The protein interaction dataset and evaluation resource. bioRxiv [Internet]. 2024 Aug 13 [cited 2025 May 20];2024.07.17.603980. Available from: https://www.biorxiv.org/content/10.1101/2024.07.17.603980v4

112. Wee JJ, Wei GW. Evaluation of AlphaFold 3's Protein-Protein Complexes for Predicting Binding Free Energy Changes upon Mutation. J Chem Inf Model [Internet]. 2024 Aug 26 [cited 2025 Jun 12];64(16):6676–83. Available from: https://pubs.acs.org/doi/full/10.1021/acs.jcim.4c00976

113. Varga JK, Ovchinnikov S, Schueler-Furman O. actifpTM: a refined confidence metric of AlphaFold2 predictions involving flexible regions. Bioinformatics [Internet]. 2025 Mar 4 [cited 2025 Jun 12];41(3). Available from: https://dx.doi.org/10.1093/bioinformatics/btaf107

114. Dunbrack RL. Rēs ipSAE loquunt: What's wrong with AlphaFold's ipTM score and how to fix it. bioRxiv [Internet]. 2025 Feb 14 [cited 2025 Jun 12];2025.02.10.637595. Available from: https://www.biorxiv.org/content/10.1101/2025.02.10.637595v1

115. Li G, Pahari S, Murthy AK, Liang S, Fragoza R, Yu H, et al. SAAMBE-SEQ: a sequence-based method for predicting mutation effect on protein–protein binding affinity. Bioinformatics [Internet]. 2021 May 17 [cited 2024 Aug 2];37(7):992–9. Available from: https://dx.doi.org/10.1093/bioinformatics/btaa761

116. Jemimah S, Sekijima M, Gromiha MM. ProAffiMuSeq: sequence-based method to predict the binding free energy change of protein–protein complexes upon mutation using functional classification. Bioinformatics [Internet]. 2020 Mar 15 [cited 2024 Aug 2];36(6):1725–30. Available from: https://dx.doi.org/10.1093/bioinformatics/btz829

117. Nikam R, Jemimah S, Gromiha MM. DeepPPAPredMut: deep ensemble method for predicting the binding affinity change in protein–protein complexes upon mutation. Bioinformatics [Internet]. 2024 May 2 [cited 2024 Aug 2];40(5). Available from: https://dx.doi.org/10.1093/bioinformatics/btae309

118. Alam N, Goldstein O, Xia B, Porter KA, Kozakov D, Schueler-Furman O. High-resolution global peptide-protein docking using fragments-based PIPER-FlexPepDock. PLoS Comput Biol [Internet]. 2017 Dec 1 [cited 2024 Aug 2];13(12):e1005905. Available from: https://journals.plos.org/ploscompbiol/article?id=10.1371/journal.pcbi.1005905

119. Moal IH, Fernández-Recio J. SKEMPI: a Structural Kinetic and Energetic database of Mutant Protein Interactions and its use in empirical models. Bioinformatics [Internet]. 2012 Oct 15 [cited 2022 Sep 1];28(20):2600–7. Available from: https://academic.oup.com/bioinformatics/article/28/20/2600/202485

120. Kumar MDS, Gromiha MM. PINT: Protein–protein Interactions Thermodynamic Database. Nucleic Acids Res [Internet]. 2006 Jan 1 [cited 2024 Jul 31];34(Database issue):D195. Available from: /pmc/articles/PMC1347380/

121. Thorn KS, Bogan AA. ASEdb: a database of alanine mutations and their effects on the free energy of binding in protein interactions. Bioinformatics [Internet]. 2001 [cited 2024 Jul 24];17(3):284–5. Available from: https://pubmed.ncbi.nlm.nih.gov/11294795/

122. Fleishman SJ, Baker D. Role of the biomolecular energy gap in protein design, structure, and evolution. Cell [Internet]. 2012 Apr 13 [cited 2023 Jul 11];149(2):262–73. Available from: http://www.cell.com/article/S0092867412003492/fulltext

123. Frenz B, Lewis SM, King I, DiMaio F, Park H, Song Y. Prediction of Protein Mutational Free Energy: Benchmark and Sampling Improvements Increase Classification Accuracy. Front Bioeng Biotechnol [Internet]. 2020 Oct 8 [cited 2022 Aug 31];8. Available from: /pmc/articles/PMC7579412/

124. Alford RF, Leaver-Fay A, Jeliazkov JR, O'Meara MJ, DiMaio FP, Park H, et al. The Rosetta All-Atom Energy Function for Macromolecular Modeling and Design. J Chem Theory Comput [Internet]. 2017 Jun 13 [cited 2024 Jan 18];13(6):3031–48. Available from: https://pubs.acs.org/doi/full/10.1021/acs.jctc.7b00125

125. Leman JK, Weitzner BD, Lewis SM, Adolf-Bryfogle J, Alam N, Alford RF, et al. Macromolecular modeling and design in Rosetta: recent methods and frameworks. Nature Methods 2020 17:7 [Internet]. 2020 Jun 1 [cited 2023 Jan 18];17(7):665–80. Available from: https://www.nature.com/articles/s41592-020-0848-2

126. Nikam R, Kulandaisamy A, Harini K, Sharma D, Michael Gromiha M. ProThermDB: thermodynamic database for proteins and mutants revisited after 15 years. Nucleic Acids Res [Internet]. 2021 Jan 8 [cited 2024 Jul 31];49(D1):D420–4. Available from: https://dx.doi.org/10.1093/nar/gkaa1035

127. Li M, Simonetti FL, Goncearenco A, Panchenko AR. MutaBind estimates and interprets the effects of sequence variants on protein–protein interactions. Nucleic Acids Res [Internet]. 2016 Jul 8 [cited 2024 Jul 30];44(W1):W494–501. Available from: https://dx.doi.org/10.1093/nar/gkw374

128. Frazer J, Notin P, Dias M, Gomez A, Min JK, Brock K, et al. Disease variant prediction with deep generative models of evolutionary data. Nature 2021 599:7883 [Internet]. 2021 Oct 27 [cited 2024 Jul 25];599(7883):91–5. Available from: https://www.nature.com/articles/s41586-021-04043-8

129. Laine E, Karami Y, Carbone A. GEMME: A Simple and Fast Global Epistatic Model Predicting Mutational Effects. Mol Biol Evol [Internet]. 2019 Nov 1 [cited 2024 Jul 25];36(11):2604–19. Available from: https://dx.doi.org/10.1093/molbev/msz179

130. Høie MH, Cagiada M, Beck Frederiksen AH, Stein A, Lindorff-Larsen K. Predicting and interpreting large-scale mutagenesis data using analyses of protein stability and conservation. Cell Rep. 2022 Jan 11;38(2):110207.

131. Cagiada M, Bottaro S, Lindemose S, Schenstrøm SM, Stein A, Hartmann-Petersen R, et al. Discovering functionally important sites in proteins. Nature Communications 2023 14:1 [Internet]. 2023 Jul 13 [cited 2024 Jul 25];14(1):1–13. Available from: https://www.nature.com/articles/s41467-023-39909-0

132. Lin Z, Akin H, Rao R, Hie B, Zhu Z, Lu W, et al. Evolutionary-scale prediction of atomic-level protein structure with a language model. Science (1979) [Internet]. 2023 Mar 17 [cited 2024 Jul 25];379(6637):1123–30. Available from: https://www.science.org/doi/10.1126/science.ade2574

133. Dehouck Y, Grosfils A, Folch B, Gilis D, Bogaerts P, Rooman M. Fast and accurate predictions of protein stability changes upon mutations using statistical potentials and neural networks: PoPMuSiC-2.0. Bioinformatics [Internet]. 2009 [cited 2022 Aug 31];25(19):2537–43. Available from: https://pubmed.ncbi.nlm.nih.gov/19654118/

134. Dieckhaus H, Brocidiacono M, Randolph NZ, Kuhlman B. Transfer learning to leverage larger datasets for improved prediction of protein stability changes. Proc Natl Acad Sci U S A [Internet]. 2024 Feb 6 [cited 2024 Jul 30];121(6):e2314853121. Available from: https://www.pnas.org/doi/abs/10.1073/pnas.2314853121

135. Pires DEV, Ascher DB, Blundell TL. mCSM: predicting the effects of mutations in proteins using graph-based signatures. Bioinformatics [Internet]. 2014 Feb 2 [cited 2022 Aug 31];30(3):335. Available from: /pmc/articles/PMC3904523/

136. Jankauskaite J, Jiménez-García B, Dapkunas J, Fernández-Recio J, Moal IH. SKEMPI 2.0: an updated benchmark of changes in protein–protein binding energy, kinetics and thermodynamics upon mutation. Bioinformatics [Internet]. 2019 Feb 2 [cited 2022 Sep 1];35(3):462. Available from: /pmc/articles/PMC6361233/

137. Geng C, Vangone A, Folkers GE, Xue LC, Bonvin AMJJ. iSEE: Interface structure, evolution, and energy-based machine learning predictor of binding affinity changes upon mutations. Proteins [Internet]. 2019 Feb 1 [cited 2024 Jul 24];87(2):110. Available from: /pmc/articles/PMC6587874/

138. Dias R, Kolaczkowski B. Improving the accuracy of high-throughput protein-protein affinity prediction may require better training data. BMC Bioinformatics [Internet]. 2017 Mar 23 [cited 2024 Jul 25];18(5):7–18. Available from: https://bmcbioinformatics.biomedcentral.com/articles/10.1186/s12859-017-1533-z

139. Tsishyn M, Pucci F, Rooman M. Quantification of biases in predictions of protein–protein binding affinity changes upon mutations. Brief Bioinform [Internet]. 2023 Nov 22 [cited 2024 Apr 4];25(1):1–11. Available from: https://dx.doi.org/10.1093/bib/bbad491

140. Valanciute A, Nygaard L, Zschach H, Jepsen MM, Lindorff-Larsen K, Stein A. Accurate protein stability predictions from homology models. bioRxiv [Internet]. 2022 Jul 13 [cited 2022 Aug 31];2022.07.12.499700. Available from: https://www.biorxiv.org/content/10.1101/2022.07.12.499700v1

141. Dobson C, Šali A, Edition MKCI, 1998 undefined. Protein folding: a perspective from theory and experiment. Angewandte Chemie (International Ed in English) [Internet]. 1998 [cited 2024 Aug 3];(37):868–93. Available from: https://onlinelibrary.wiley.com/doi/abs/10.1002/(SICI)1521-3773(19980420)37:7%3C868::AID-ANIE868%3E3.0.CO;2-H

142. Shoichet BK, Baase WA, Kuroki R, Matthews BW. A relationship between protein stability and protein function. Proc Natl Acad Sci U S A [Internet]. 1995 Jan 17 [cited 2024 Aug 3];92(2):452–6. Available from: https://www.pnas.org

143. Tokuriki N, Stricher F, Schymkowitz J, Serrano L, Tawfik DS. The Stability Effects of Protein Mutations Appear to be Universally Distributed. J Mol Biol. 2007 Jun 22;369(5):1318–32.

144. Heyne M, Papo N, Shifman JM. Generating quantitative binding landscapes through fractional binding selections combined with deep sequencing and data normalization. Nature Communications 2020 11:1 [Internet]. 2020 Jan 15 [cited 2024 Jul 31];11(1):1–7. Available from: https://www.nature.com/articles/s41467-019-13895-8

145. Coulbourn S, Id F, Id AA, Glaros Id A. Mining the Protein Data Bank to improve prediction of changes in protein-protein binding. Helmer-Citterich M, editor. PLoS One [Internet]. 2021 Nov 2 [cited 2021 Dec 6];16(11):e0257614. Available from: https://journals.plos.org/plosone/article?id=10.1371/journal.pone.0257614

146. Baker D. An exciting but challenging road ahead for computational enzyme design. Protein Science [Internet]. 2010 Oct 1 [cited 2025 Jun 25];19(10):1817–9. Available from: https://onlinelibrary.wiley.com/doi/full/10.1002/pro.481

147. Li Y, Li H, Yang F, Smith-Gill SJ, Mariuzza RA. X-ray snapshots of the maturation of an antibody response to a protein antigen. Nature Structural & Molecular Biology 2003 10:6 [Internet]. 2003 May 12 [cited 2025 Jun 26];10(6):482–8. Available from: https://www.nature.com/articles/nsb930

148. Smith CA, Kortemme T. Backrub-Like Backbone Simulation Recapitulates Natural Protein Conformational Variability and Improves Mutant Side-Chain Prediction. J Mol Biol. 2008 Jul 18;380(4):742–56.

149. Ollikainen N, Jong RM de, Kortemme T. Coupling Protein Side-Chain and Backbone Flexibility Improves the Re-design of Protein-Ligand Specificity. PLoS Comput Biol [Internet]. 2015 [cited 2025 May 20];11(9):e1004335. Available from: https://journals.plos.org/ploscompbiol/article?id=10.1371/journal.pcbi.1004335

150. Kuroda D, Gray JJ. Pushing the backbone in protein-protein docking. Structure [Internet]. 2016 Oct 4 [cited 2025 Jun 26];24(10):1821. Available from: https://pmc.ncbi.nlm.nih.gov/articles/PMC5069389/

151. Hunter SA, Cochran JR. Cell-Binding Assays for Determining the Affinity of Protein–Protein Interactions: Technologies and Considerations. Methods Enzymol. 2016 Jan 1;580:21–44.

152. Gray VE, Hause RJ, Luebeck J, Shendure J, Fowler DM. Quantitative Missense Variant Effect Prediction Using Large-Scale Mutagenesis Data. Cell Syst [Internet]. 2018 Jan 24 [cited 2024 Aug 3];6(1):116-124.e3. Available from: http://www.cell.com/article/S2405471217304921/fulltext

153. Perica T, Mathy CJP, Xu J, Jang G, Zhang Y, Kaake R, et al. Systems-level effects of allosteric perturbations to a model molecular switch. Nature 2021 599:7883 [Internet]. 2021 Oct 13 [cited 2024 Aug 5];599(7883):152–7. Available from: https://www.nature.com/articles/s41586-021-03982-6

154. Yu D, Chojnowski G, Rosenthal M, Kosinski J. AlphaPulldown—a python package for protein–protein interaction screens using AlphaFold-Multimer. Bioinformatics [Internet]. 2023 Jan 1 [cited 2024 Aug 5];39(1). Available from: https://dx.doi.org/10.1093/bioinformatics/btac749

155. Lu H, Zhou Q, He J, Jiang Z, Peng C, Tong R, et al. Recent advances in the development of protein–protein interactions modulators: mechanisms and clinical trials. Signal Transduction and Targeted Therapy 2020 5:1 [Internet]. 2020 Sep 23

[cited 2024 Aug 5];5(1):1–23. Available from: https://www.nature.com/articles/s41392-020-00315-3